\pdfoutput=1
\documentclass[12pt,a4paper]{article}
\usepackage{ifthen} % for conditional statements
\newboolean{pdflatex}
\setboolean{pdflatex}{true} % False for eps figures
\usepackage{siunitx}
\newboolean{articletitles}
\setboolean{articletitles}{true} % False removes titles in references

\newboolean{uprightparticles}
\setboolean{uprightparticles}{false} %True for upright particle symbols

\newboolean{inbibliography}
\setboolean{inbibliography}{false} %True once you enter the bibliography

\def\paperauthors{LHCb collaboration} % Leave as is for PAPER, CONF and FIGURE
\def\paperasciititle{} % Set ASCII title here !! MAKE sure it's only ASCII characters !!
\def\papercopyright{} % Set ASCII title here !! MAKE sure it's only ASCII characters !!
\def\papertitle{First observation of the charmless baryonic decay $\BuLPbarPPbar$}

\def\paperkeywords{{High Energy Physics}, {LHCb}} % Comma separated list
\def\papercopyright{CERN on behalf of the LHCb collaboration}
\def\paperlicence{CC BY 4.0 licence}
\def\paperlicenceurl{https://creativecommons.org/licenses/by/4.0/}

\usepackage[top=1in, bottom=1.25in, left=1in, right=1in]{geometry}

\usepackage{microtype}
\usepackage{lineno}  % for line numbering during review
\usepackage{xspace} % To avoid problems with missing or double spaces after
\usepackage{caption} %these three command get the figure and table captions automatically small

\usepackage{graphicx}  % to include figures (can also use other packages)
\usepackage{color}
\usepackage{colortbl}
\graphicspath{{./figs/}} % Make Latex search fig subdir for figures
\usepackage{amsmath} % Adds a large collection of math symbols
\usepackage{amssymb}
\usepackage{amsfonts}
\usepackage{upgreek} % Adds in support for greek letters in roman typeset

\newcommand*\patchAmsMathEnvironmentForLineno[1]{%
\expandafter\let\csname old#1\expandafter\endcsname\csname #1\endcsname
\expandafter\let\csname oldend#1\expandafter\endcsname\csname
end#1\endcsname
 \renewenvironment{#1}%
   {\linenomath\csname old#1\endcsname}%
   {\csname oldend#1\endcsname\endlinenomath}%
}
\newcommand*\patchBothAmsMathEnvironmentsForLineno[1]{%
  \patchAmsMathEnvironmentForLineno{#1}%
  \patchAmsMathEnvironmentForLineno{#1*}%
}
\AtBeginDocument{%
\patchBothAmsMathEnvironmentsForLineno{equation}%
\patchBothAmsMathEnvironmentsForLineno{align}%
\patchBothAmsMathEnvironmentsForLineno{flalign}%
\patchBothAmsMathEnvironmentsForLineno{alignat}%
\patchBothAmsMathEnvironmentsForLineno{gather}%
\patchBothAmsMathEnvironmentsForLineno{multline}%
\patchBothAmsMathEnvironmentsForLineno{eqnarray}%
}

\usepackage[pdftex,
            pdfauthor={\paperauthors},
            pdftitle={\paperasciititle},
            pdfkeywords={\paperkeywords},
            ]{hyperref}
\usepackage{hyperxmp}

\usepackage[colorinlistoftodos,textsize=scriptsize]{todonotes}

\usepackage[bottom,flushmargin,hang,multiple]{footmisc}

\usepackage[all]{hypcap} % Internal hyperlinks to floats.

\usepackage{xspace} 
\usepackage{upgreek}

\def\lhcb   {\mbox{LHCb}\xspace}

\def\MagUp {\mbox{\em Mag\kern -0.05em Up}\xspace}

\ifthenelse{\boolean{uprightparticles}}%
{

 \def\Peta        {\ensuremath{\upeta}\xspace}

 \def\Ppi         {\ensuremath{\uppi}\xspace}

 \def\Ppsi        {\ensuremath{\uppsi}\xspace}

 \def\PDelta      {\ensuremath{\Delta}\xspace}                 
 \def\PXi         {\ensuremath{\Xi}\xspace}                 
 \def\PLambda     {\ensuremath{\Lambda}\xspace}                 
 \def\PSigma      {\ensuremath{\Sigma}\xspace}                 
 \def\POmega      {\ensuremath{\Omega}\xspace}                 
 \def\PUpsilon    {\ensuremath{\Upsilon}\xspace}
 \let\oldPi\Pi
 \def\PPi         {\ensuremath{\oldPi}\xspace}

 \def\PB      {\ensuremath{\mathrm{B}}\xspace}                 
                  
 \def\PD      {\ensuremath{\mathrm{D}}\xspace}

 \def\PJ      {\ensuremath{\mathrm{J}}\xspace}                 
 \def\PK      {\ensuremath{\mathrm{K}}\xspace}

 \def\Pb      {\ensuremath{\mathrm{b}}\xspace}                 
 \def\Pc      {\ensuremath{\mathrm{c}}\xspace}

 \def\Pi      {\ensuremath{\mathrm{i}}\xspace}

 \def\Pp      {\ensuremath{\mathrm{p}}\xspace}

 \def\Ps      {\ensuremath{\mathrm{s}}\xspace}

 \def\thebaroffset{0.0em}
}
{

 \def\Peta        {\ensuremath{\eta}\xspace}

 \def\Ppi         {\ensuremath{\pi}\xspace}

 \def\Ppsi        {\ensuremath{\psi}\xspace}                 
                  
 \mathchardef\PDelta="7101
 \mathchardef\PXi="7104
 \mathchardef\PLambda="7103
 \mathchardef\PSigma="7106
 \mathchardef\POmega="710A
 \mathchardef\PUpsilon="7107
 \mathchardef\PPi="7105
                  
 \def\PB      {\ensuremath{B}\xspace}                 
                  
 \def\PD      {\ensuremath{D}\xspace}

 \def\PJ      {\ensuremath{J}\xspace}                 
 \def\PK      {\ensuremath{K}\xspace}

 \def\Pb      {\ensuremath{b}\xspace}                 
 \def\Pc      {\ensuremath{c}\xspace}

 \def\Pi      {\ensuremath{i}\xspace}

 \def\Pp      {\ensuremath{p}\xspace}

 \def\Ps      {\ensuremath{s}\xspace}

 \def\thebaroffset{0.18em}
}
\newcommand{\offsetoverline}[2][\thebaroffset]{\kern #1\overline{\kern -#1 #2}}%

\makeatletter
\ifcase \@ptsize \relax% 10pt
  \newcommand{\miniscule}{\@setfontsize\miniscule{4}{5}}% \tiny: 5/6
\or% 11pt
  \newcommand{\miniscule}{\@setfontsize\miniscule{5}{6}}% \tiny: 6/7
\or% 12pt
  \newcommand{\miniscule}{\@setfontsize\miniscule{5}{6}}% \tiny: 6/7
\fi
\makeatother

\DeclareRobustCommand{\optbar}[1]{\shortstack{{\miniscule (\rule[.5ex]{1.25em}{.18mm})}
  \\ [-.7ex] $#1$}}

\def\squark    {{\ensuremath{\Ps}}\xspace}

\def\cquark    {{\ensuremath{\Pc}}\xspace}
\def\cquarkbar {{\ensuremath{\overline \cquark}}\xspace}
\def\ccbar     {{\ensuremath{\cquark\cquarkbar}}\xspace}
\def\bquark    {{\ensuremath{\Pb}}\xspace}

\def\pion   {{\ensuremath{\Ppi}}\xspace}

\def\pip    {{\ensuremath{\pion^+}}\xspace}
\def\pim    {{\ensuremath{\pion^-}}\xspace}

\def\kaon    {{\ensuremath{\PK}}\xspace}
\def\KorKbar {\kern \thebaroffset\optbar{\kern -\thebaroffset \PK}{}\xspace}

\def\Kp      {{\ensuremath{\kaon^+}}\xspace}
\def\Km      {{\ensuremath{\kaon^-}}\xspace}

\def\KS      {{\ensuremath{\kaon^0_{\mathrm{S}}}}\xspace}

\def\D       {{\ensuremath{\PD}}\xspace}

\def\DorDbar {\kern \thebaroffset\optbar{\kern -\thebaroffset \PD}\xspace}

\def\Dp      {{\ensuremath{\D^+}}\xspace}
\def\Dm      {{\ensuremath{\D^-}}\xspace}

\def\DpDm    {\ensuremath{\Dp {\kern -0.16em \Dm}}\xspace}

\def\B       {{\ensuremath{\PB}}\xspace}

\def\BorBbar {\kern \thebaroffset\optbar{\kern -\thebaroffset \PB}\xspace}
\def\Bz      {{\ensuremath{\B^0}}\xspace}

\def\Bd      {{\ensuremath{\B^0}}\xspace}

\def\BdorBdbar {\kern \thebaroffset\optbar{\kern -\thebaroffset \Bd}\xspace}
\def\Bu      {{\ensuremath{\B^+}}\xspace}
\def\Bub     {{\ensuremath{\B^-}}\xspace}
\def\Bp      {{\ensuremath{\Bu}}\xspace}
\def\Bm      {{\ensuremath{\Bub}}\xspace}

\def\Bs      {{\ensuremath{\B^0_\squark}}\xspace}

\def\BsorBsbar {\kern \thebaroffset\optbar{\kern -\thebaroffset \Bs}\xspace}

\def\Bds     {{\ensuremath{\B_{(\squark)}^0}}\xspace}

\def\jpsi     {{\ensuremath{{\PJ\mskip -3mu/\mskip -2mu\Ppsi}}}\xspace}

\def\etac     {{\ensuremath{\Peta_\cquark}}\xspace}

\def\Y#1S{\ensuremath{\PUpsilon{(#1S)}}\xspace}

\def\proton      {{\ensuremath{\Pp}}\xspace}
\def\antiproton  {{\ensuremath{\overline \proton}}\xspace}

\def\Lz          {{\ensuremath{\PLambda}}\xspace}
\def\Lbar        {{\ensuremath{\offsetoverline{\PLambda}}}\xspace}
\def\LorLbar     {\kern \thebaroffset\optbar{\kern -\thebaroffset \PLambda}\xspace}

\def\BF         {{\ensuremath{\mathcal{B}}}\xspace}

\newcommand{\decay}[2]{\ensuremath{#1\!\to #2}\xspace} 

\def\to                 {\ensuremath{\rightarrow}\xspace}

\def\CP                {{\ensuremath{C\!P}}\xspace}

\def\AT#1     {\ensuremath{A_{\mathrm{T}}^{#1}}\xspace}           % 2

\def\C#1      {\ensuremath{\mathcal{C}_{#1}}\xspace}                       % 9
\def\Cp#1     {\ensuremath{\mathcal{C}_{#1}^{'}}\xspace}                    % 7
\def\Ceff#1   {\ensuremath{\mathcal{C}_{#1}^{\mathrm{(eff)}}}\xspace}        % 9  
\def\Cpeff#1  {\ensuremath{\mathcal{C}_{#1}^{'\mathrm{(eff)}}}\xspace}       % 7
\def\Ope#1    {\ensuremath{\mathcal{O}_{#1}}\xspace}                       % 2
\def\Opep#1   {\ensuremath{\mathcal{O}_{#1}^{'}}\xspace}                    % 7

\newcommand{\aunit}[1]{\ensuremath{\text{\,#1}}}       
\newcommand{\tev}{\aunit{Te\kern -0.1em V}\xspace}
\newcommand{\gev}{\aunit{Ge\kern -0.1em V}\xspace}
\newcommand{\mev}{\aunit{Me\kern -0.1em V}\xspace}
\newcommand{\kev}{\aunit{ke\kern -0.1em V}\xspace}
\newcommand{\ev}{\aunit{e\kern -0.1em V}\xspace}
 
\newcommand{\mevc}{\ensuremath{\aunit{Me\kern -0.1em V\!/}c}\xspace}
\newcommand{\gevc}{\ensuremath{\aunit{Ge\kern -0.1em V\!/}c}\xspace}
\newcommand{\mevcc}{\ensuremath{\aunit{Me\kern -0.1em V\!/}c^2}\xspace}
\newcommand{\gevcc}{\ensuremath{\aunit{Ge\kern -0.1em V\!/}c^2}\xspace}
\def\fb   {\ensuremath{\aunit{fb}}\xspace}
\def\invfb   {\ensuremath{\fb^{-1}}\xspace}

\newcommand{\chisq}{\ensuremath{\chi^2}\xspace}

\newcommand{\chisqip}{\ensuremath{\chi^2_{\text{IP}}}\xspace}

\def\gsim{{~\raise.15em\hbox{$>$}\kern-.85em
          \lower.35em\hbox{$\sim$}~}\xspace}
\def\lsim{{~\raise.15em\hbox{$<$}\kern-.85em
          \lower.35em\hbox{$\sim$}~}\xspace}

\def\sPlot{\mbox{\em sPlot}\xspace}

\def\pt         {\ensuremath{p_{\mathrm{T}}}\xspace}

\def\evtgen     {\mbox{\textsc{EvtGen}}\xspace}

\def\geant      {\mbox{\textsc{Geant4}}\xspace}

\def\photos     {\mbox{\textsc{Photos}}\xspace}

\def\pythia     {\mbox{\textsc{Pythia}}\xspace}

\def\tell1  {TELL1\xspace}
\def\ukl1   {UKL1\xspace}

\newcommand{\lhcborcid}[1]{\href{https://orcid.org/#1}{\hspace*{0.1em}\raisebox{-0.45ex}{\includegraphics[width=1em]{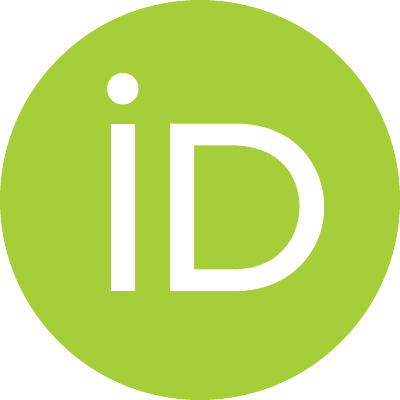}}}}

\usepackage{cite} % Allows for ranges in citations
\usepackage{mciteplus}
\newcommand{\xx}{\ensuremath{\kern 0.5em }}

\def\pp {\ensuremath{pp}\xspace}

\def\sPlot{\mbox{\em sPlot}\xspace}

\usepackage{multirow} % for complicated table
\usepackage{booktabs} % for complicated table
\usepackage{rotating}

\def\Bds{{\ensuremath{\B^0_{\kern -0.1em{\scriptscriptstyle (}\kern -0.05em\squark\kern -0.03em{\scriptscriptstyle )}}}}\xspace}

\def\normalization{\mbox{\decay{\Bp}{{\jpsi}(\to{\Lbar\proton\Km}){\Kp}}}}
\def\normalizationcc{\decay{\Bp}{{\jpsi}(\to{\Lbar\proton\Km+c.c.}){\Kp}}}
\newcommand{\BuLPbarPPbar}{\texorpdfstring{\decay{\Bp}{\Lbar\proton \antiproton \proton}}{}}
\newcommand{\BuLPbarEtac}{\texorpdfstring{\decay{\Bp}{\Lbar\proton \etac(\to \antiproton \proton)}}{}}

\usepackage{longtable} % only for template; not usually to be used in PAPERs
\usepackage{subcaption}

\usepackage[normalem]{ulem}

\begin{document}

%%%%%%%%%%%%%%%%%%%%%%%%%
%%%%% Title     %%%%%%%%%
%%%%%%%%%%%%%%%%%%%%%%%%%
\renewcommand{\thefootnote}{\fnsymbol{footnote}}
\setcounter{footnote}{1}

% %%%%%%% CHOOSE TITLE PAGE--------
%\onecolumn
% TC:ignore
% ===============================================================================
% Purpose: LHCb-PAPER journal paper title page template
% Author: 
% Created on: 2010-09-25
% ===============================================================================

%%%%%%%%%%%%%%%%%%%%%%%%%
%%%%%  TITLE PAGE  %%%%%%
%%%%%%%%%%%%%%%%%%%%%%%%%
\begin{titlepage}
\pagenumbering{roman}

% Header ---------------------------------------------------
\vspace*{-1.5cm}
\centerline{\large EUROPEAN ORGANIZATION FOR NUCLEAR RESEARCH (CERN)}
\vspace*{1.5cm}
\noindent
\begin{tabular*}{\linewidth}{lc@{\extracolsep{\fill}}r@{\extracolsep{0pt}}}
\ifthenelse{\boolean{pdflatex}}% Logo format choice
{\vspace*{-1.5cm}\mbox{\!\!\!\includegraphics[width=.14\textwidth]{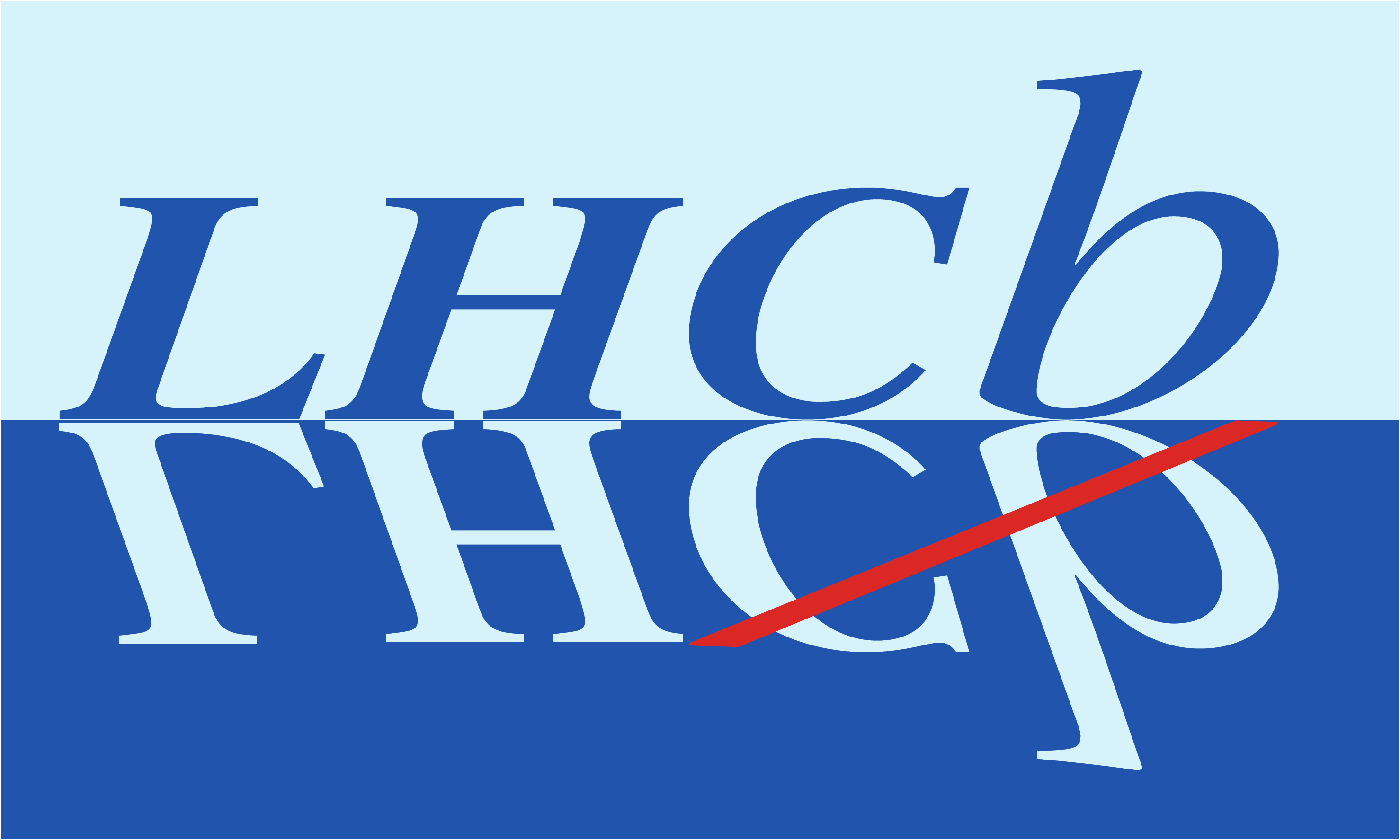}} & &}%
{\vspace*{-1.2cm}\mbox{\!\!\!\includegraphics[width=.12\textwidth]{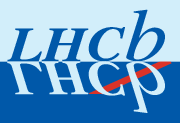}} & &}%
\\
 & & CERN-EP-2025-174 \\  % ID 
 & & LHCb-PAPER-2025-032 \\  % ID 
 & & \today \\ % Date - Can also hardwire e.g.: 23 March 2010
 & & \\
\end{tabular*}

\vspace*{2.5cm}

% Title --------------------------------------------------
{\normalfont\bfseries\boldmath\huge
\begin{center}
% DO NOT EDIT HERE. Instead edit macro in main.tex to keep metadata correct
  \papertitle 
\end{center}
}

\vspace*{2.0cm}

% Authors -------------------------------------------------
\begin{center}
\paperauthors
\footnote{Full author list given at the end of the article.}
%In the footnote, replace 'paper' by 'Letter' in case of submission to PRL or PLB 
 \end{center}

\vspace{\fill}

% Abstract -----------------------------------------------
\begin{abstract}
  \noindent
A search for the charmless baryonic decay $B^+\to \Lbar p\bar{p}p$ is performed using proton-proton collision data recorded by the LHCb experiment, corresponding to an integrated luminosity of 5.4\invfb.
The branching fraction for this decay is measured for the first time relative to that of the topologically similar decay \decay{\Bp}{\jpsi\Kp}, with \decay{\jpsi}{\Lbar\proton\Km}.
  The branching fraction is measured to be
\mbox{$\mathcal{B}(\Bp \to \Lbar\proton\antiproton\proton) = (2.15\pm0.35\pm0.12\pm0.28) \times 10^{-7}$},
where the first uncertainty is statistical, the second is systematic, and the third arises from the uncertainty in the normalization channel
branching fraction.
  The \CP asymmetry is measured to be $\mathcal{A}_{\CP}=(5.4\pm 15.6\pm 2.4)\%$, where the uncertainties are statistical and systematic. 
 The background-subtracted invariant-mass distributions of $\Lbar\proton$ and $\antiproton\proton$ pairs exhibit pronounced enhancements at both kinematic thresholds, in contrast to a uniform phase-space distribution. 
\end{abstract}

\begin{center}
  Published in Physical Review Letters \textbf{135} (2025) 261901  
  % Submitted to Phys.~Rev.~Lett.
\end{center}

\vspace{\fill}

{\footnotesize 
% Edit macro in main.tex to keep metadata correct
\centerline{\copyright~\papercopyright. \href{\paperlicenceurl}{\paperlicence}.}}
\vspace*{2mm}

\end{titlepage}

%%%%%%%%%%%%%%%%%%%%%%%%%%%%%%%%
%%%%%  EOD OF TITLE PAGE  %%%%%%
%%%%%%%%%%%%%%%%%%%%%%%%%%%%%%%%

%  empty page follows the title page ----
\newpage
\setcounter{page}{2}

% TC:endignore
% \twocolumn
% %%%%%%%%%%%%% ---------

\renewcommand{\thefootnote}{\arabic{footnote}}
\setcounter{footnote}{0}

%%%%%%%%%%%%%%%%%%%%%%%%%%%%%%%%
%%%%%  Table of Content   %%%%%%
%%%%%%%%%%%%%%%%%%%%%%%%%%%%%%%%
%%%% Uncomment if desired
% \tableofcontents
\cleardoublepage

%%%%%%%%%%%%%%%%%%%%%%%%%
%%%%% Main text %%%%%%%%%
%%%%%%%%%%%%%%%%%%%%%%%%%

\pagestyle{plain} % restore page numbers for the main text
\setcounter{page}{1}
\pagenumbering{arabic}
%% Uncomment during review phase.
%% Comment before a final submission.
% \linenumbers
%% ================
%%   Introduction
%% ================
Studies of exclusive $B$-meson decays to final states containing both baryons and mesons have been ongoing since the late 1980s~\cite{ARGUS:1988gzx}. Several baryonic $B$ decay modes have been observed, including three-body final states such as $\proton\antiproton\Kp$, $\proton\antiproton\pip$, $\proton\antiproton\KS$, and $\proton\antiproton K^{*}(892)^{+}$~\cite{Belle:2002bro,Belle:2003pwf}, as well as four-body modes like $\proton\antiproton\pip\pim$ and $\proton\antiproton\Kp\Km$~\cite{LHCb-PAPER-2017-005}. In the Standard Model, these decays proceed via $b \to u$ tree-level and $b \to s(d)$ electroweak penguin diagrams, providing a valuable laboratory for studying direct \CP violation and testing hadronic models~\cite{Cheng:2001tr, Chua:2002wn}.

Unlike mesonic modes, baryonic \B decays exhibit a unique feature known as threshold enhancement: a pronounced peak near the invariant-mass threshold of the baryon–antibaryon pair~\cite{Belle:2002bro,Belle:2003pwf}. Final states with two baryon–antibaryon pairs are particularly intriguing, as they probe hadronization dynamics beyond those accessible with a single pair~\cite{Hsiao:2022tfj}.

The \lhcb collaboration recently reported the first observation of a four-body charmless fully baryonic weak decay, $\Bz \to \proton\antiproton \proton\antiproton$~\cite{LHCb-PAPER-2022-032}, which mainly proceeds via a $b \to u$ tree-level internal $W$-boson emission~\cite{Hsiao:2022tfj}. The measured branching fraction, \mbox{$\mathcal{B}(\decay{\Bz}{\proton\antiproton \proton\antiproton})=(2.2\pm0.4\pm0.1\pm0.1)\times10^{-8}$}, is approximately two orders of magnitude lower than those of analogous four-body baryonic \B decays, such as $\Bz  \to \proton\antiproton\pip\pim$~\cite{LHCb-PAPER-2017-005}. To explain the overall observed suppression, form factors that describe the evolution of short-distance systems to those with baryons in the final state have been
proposed~\cite{Hsiao:2022tfj}.
In this approach, a \emph{double-enhancement} mechanism in baryon-pair formation was also proposed. 
It increases the decay rate in the limited regions of phase space where one or both baryon pairs are produced near threshold.
With this effect, the decay \BuLPbarPPbar,\footnote{The inclusion of charge-conjugate processes is implied throughout, except in the discussion of asymmetries.} which proceeds via a combination of  CKM-suppressed $ b \to u $ tree level and   $b \to s$ electroweak loop amplitudes, is expected to have a branching fraction of \mbox{$\mathcal{B}(\BuLPbarPPbar) = (7.4^{+0.6}_{-0.2} \pm 0.03 ^{+3.6}_{-2.6}) \times 10^{-7}$~\cite{Hsiao:2022tfj}}, where the uncertainty arises from nonfactorizable QCD corrections, the CKM matrix-element inputs, and form‐factor modeling. The comparatively large branching fraction of this decay offers a promising opportunity to study the mass spectra of baryon pairs and explore baryon-pair formation mechanisms. The mass spectra may reveal resonant structures such as the $X(1835)$ and $X(2085)$, which are interpreted as baryonium-like bound states~\cite{BESIII:2011aa,BESIII:2023kgz}. In addition, owing to the interference between tree‐level and penguin diagrams in the low $m(\Lbar\proton)$ and $m(\antiproton\proton)$ mass regions, sizeable direct \CP asymmetries are possible~\cite{Geng:2006jt}.

In this Letter, the \BuLPbarPPbar decay is studied using proton-proton (\pp) collision data collected by the \lhcb experiment over the period 2016--2018 at a center-of-mass energy  $\sqrt{s}=$13\tev and corresponding to an integrated luminosity of approximately 5.4\invfb. The branching fraction of the signal decay is measured relative to that of the topologically similar normalization mode, \normalization, using a simultaneous fit across all data-taking periods. In addition, this work presents a study of the $m(\Lbar\proton)$ and $m(\antiproton\proton)$ spectra and \CP asymmetry in a purely baryonic four-body decay.

%% ====================================
%%   The LHCb detector and data samples
%% ====================================
The \lhcb detector~\cite{LHCb-DP-2008-001,LHCb-DP-2014-002} is a single-arm forward spectrometer covering the pseudorapidity range $2<\eta<5$, designed for the study of particles containing \bquark or \cquark quarks. The detector used for this analysis includes a high-precision tracking system consisting of a silicon-strip vertex detector (VELO) surrounding the \pp interaction region~\cite{LHCb-DP-2014-001}, a large-area silicon-strip detector located upstream of a dipole magnet with a bending power of about $4{\mathrm{\,T\,m}}$, and three stations of silicon-strip detectors and straw drift tubes~\cite{LHCb-DP-2017-001} placed downstream of the magnet. 
The tracking system provides measurements of the track momentum and impact parameter (IP) and is used to reconstruct the primary $\pp$ interaction vertex (PV), where the IP is the shortest distance between the extrapolation of a track and a PV.
Different types of charged hadrons are distinguished using particle identification (PID) information from two ring-imaging Cherenkov detectors~\cite{LHCb-DP-2012-003}. The $\Lz\to \proton\pim$ decay is reconstructed in two distinct categories depending on whether the decay products are reconstructed in all tracking detectors (\emph{long}) or in all but the VELO (\emph{downstream}). While candidates in the long category have better mass, momentum, and vertex resolution than those in the downstream category, the downstream category accounts for two‐thirds of the total \Lz yield.

Online event selection is performed by a trigger~\cite{LHCb-DP-2012-004} that consists of a hardware stage, based on information from the calorimeter and muon systems, followed by a two-level software stage, which applies a full event reconstruction. 
At the hardware stage, events are required to have a muon with high transverse momentum (\pt) or a hadron, photon or electron with high transverse energy in the calorimeters. The first-level software trigger requires one or two tracks that are likely to originate from $B$ decays. 
The second-level software trigger requires a two-, three- or four-track secondary vertex with a significant displacement from any PV.  
A multivariate algorithm~\cite{LHCb-PROC-2015-018} is to identify secondary vertices consistent with the decay of a $b$ hadron.

Simulated samples are used to study the properties of the signal, normalization, and background channels. 
%% mds Generation of \pp collisions is performed using \pythia~\cite{Sjostrand:2007gs}, with a specific \lhcb configuration~\cite{LHCb-PROC-2010-056}. 
\pythia~\cite{Sjostrand:2007gs} generates \pp collisions using an {\lhcb}-specific configuration~\cite{LHCb-PROC-2010-056}. 
Decays of unstable particles are described by \evtgen~\cite{Lange:2001uf}, in which final-state radiation is generated using \photos~\cite{Photos-2}.
The interactions of generated particles with the detector material, and their responses, are implemented using the \geant toolkit~\cite{Agostinelli:2002hh,Allison:2006ve}, as described in Ref.~\cite{LHCb-PROC-2011-006}. 
%% Simulated signal and normalization decays are generated according to phase-space models.
Simulated signal and normalization decays are generated using phase-space models.

%% =====================
%%   Candidate selection
%% =====================
Offline candidates are selected by exploiting the characteristic topology and kinematic features of four-body decays to final states containing a $\Lz$ baryon. The preselection requires loose track quality and applies minimal kinematic requirements to the \Lz and \Bp candidates. All charged tracks except those used to reconstruct $\Lz$ candidates must originate from within the VELO and are required to have a good track-fit quality. Fake tracks, which are reconstructed from detector hits that do not correspond to real particles, are rejected by applying a selection on the fake-track probability~\cite{DeCian:2255039}. A minimum \pt of $250\mevc$, $200\mevc$, and $400\mevc$ is required for the proton, kaon, and $\Lz$ baryon, respectively. Each track must have a significant \chisqip, where \chisqip is defined as the difference between the vertex-fit $\chi^2$ of a PV reconstructed with and without the considered track.
%% mds Candidate $\Lz$ baryons are formed by combining one proton and one negative-charged pion track candidate. 
Candidate $\Lz$ baryons are formed by combining proton and pion track candidates
that are fitted to a common decay vertex. 
%% mds The two tracks are fitted to a common decay vertex.
A loose PID requirement on the proton from the $\Lz$ baryon is applied at this stage. For $\Lz$ candidates in the long category, the selection algorithm further requires that the decay vertices of the $\Lz$ baryon and the associated \Bp candidate are well separated. The $\Lz$ decay products must satisfy $|m(p\pi)-M_{\Lz}|<6\mevcc$, where $M_\Lz$ is the known $\Lz$ mass \cite{PDG2024}. Candidate \Bp mesons are reconstructed by combining a $\Lz$ candidate with three charged final-state particles into a common vertex. %Both the \BuLPbarPPbar and the \normalization~decay chains are refitted using the known $\Lz$ mass~\cite{PDG2024}. 
Both the \BuLPbarPPbar and the \normalization~decay chains are fitted constraining the $\Lz$ mass to its known value~\cite{PDG2024}. 
The PV that fits best to the flight direction of the \Bp candidate is taken as the associated PV. 
%% mds This condition is further reinforced by requiring that 
The angle between the \Bp momentum direction and the displacement vector from the PV to the \Bp decay vertex is required to be less than 2.5 degrees. 
The \Bp candidate is also required to have $\pt>3000\mevc$ and a large decay length significance, defined as the $\chisq$ of the separation between the \Bp decay vertex and the associated PV, ensuring a clear displacement from the PV.

The final selection of \Bp candidates uses the two most powerful variables that discriminate between signal and background: 
$\chisqip$ of the \Bp candidate, defined as the difference between the vertex-fit $\chi^2$ of a PV reconstructed with and without the $B^+$ meson, 
and the product of the nominal particle-identification classifiers of the three charged final-state particles, $\prod_i \mathcal{P}_i$.
% Polyakovthe product of the three charged final-state particles' nominal identification probabilities $\prod_i\mathcal{P}_{i}$.
The $\chisqip$  distribution for signal is approximately that of $ \chi^2 $ for one degree of freedom. Each classifier, $\mathcal{P}_i$, is derived from an artificial neural network and ranges between 0 and 1, peaking near unity for correctly identified particles and near zero for misidentified ones~\cite{LHCb-DP-2014-002}. 
% Each $\mathcal{P}_{i}$, $\in (0,1)$, is determined using artificial neural networks and peaks sharply near 1 for particles of type $i$ and near 0 for other species. 
Using \BuLPbarPPbar data in the sideband mass regions, defined as \mbox{$m(\Lbar\proton\antiproton\proton)\in[4700, 5049]\cup[5529,5800]\mevcc$}, along with simulated \Bp signal decays, the selection criteria are optimized separately for the long and downstream categories to enhance the signal significance ${S}/\sqrt{S+B}$, 
even if the signal is somewhat lower than that expected.
Here, $S$ and $B$ denote the expected signal and background yields within the signal region \mbox{$m(\Lbar\proton\antiproton\proton)\in[5254,5304]\mevcc$}. 
A common requirement $\log(\chisqip)<1.8~(1.6)$, along with a selection on the variable $\prod_i\mathcal{P}_{i}>0.6~(0.7)$, is applied to the long (downstream) $\Lz$ category for the signal and normalization channels. 
With respect to preselection criteria, the final selection criteria reduces background by a factor of 5 while retaining about 65\% of the signal. 
To avoid experimenter bias, data in the \BuLPbarPPbar~signal region are
examined only after all selection criteria are finalized. 

Reconstruction efficiencies for signal and normalization channels are evaluated using simulated decays generated uniformly in phase space. 
To correct the kinematic distributions in the simulated samples, candidates are assigned weights derived from background-subtracted \mbox{$\Bp\to H_{\ccbar} \, (\to \Lbar\proton\Km )\Kp$} candidates, where $H_{\ccbar}$ represents a \jpsi or \etac~meson. Weighting depends on $\chisqip$ of the \Bp candidate, track multiplicity in the event, and the momentum of each final-state particle. 
Additionally, the invariant-mass distributions of the two-body combinations of the final-state particles in simulation are corrected to match the corresponding distributions in data. 
For the normalization channel, the $m(\proton\Km)$ versus $m(\Lbar K)$ distributions are corrected, while for the signal channel, the $m(\Lbar\proton)$ versus $m(\antiproton\proton)$  distributions are corrected to better match the data after the decay is observed. 

Several $\Bp$ decays with intermediate charmonium resonances can contribute to the $\Lbar\proton\antiproton\proton$ final state, including $\Bp \to \Lbar\proton\jpsi(\to \antiproton\proton)$ and $\Bp \to \Lbar\proton\eta_c(\to \antiproton\proton)$. 
The $\Bp\to \Lbar\proton\antiproton\proton$ candidates are  divided into three disjoint categories: non-charmonium candidates with \mbox{$m(\antiproton\proton) < 2850\mevcc$}, \mbox{\BuLPbarEtac} candidates with $|m(\antiproton\proton) - M_{\etac}| < 40\mevcc$ , and \mbox{$ \Bp \to \Lbar\proton\jpsi(\to \antiproton\proton)$} candidates with $|m(\antiproton\proton) - M_{\jpsi}| < 30\mevcc$, where $M_{\etac}$ and $M_{\jpsi}$ are the known \etac and \jpsi masses~\cite{PDG2024}. The wider window for \mbox{\BuLPbarEtac} than for 
$B^+ \to \Lbar\proton\jpsi(\to \antiproton\proton)$ accounts for the larger \etac 
natural linewidth.

%% mds \footnote{The $m(\antiproton\proton)$ observed width is approximately $17.1\mevcc$ for $\eta_c\to\antiproton\proton$ and $7.5\mevcc$ for $\jpsi\to\antiproton\proton$.}

%% ===============================
%%   Signals and BF determinations
%% ===============================
\begin{figure}[htb]
    \centering
    \includegraphics[width=0.45\textwidth]{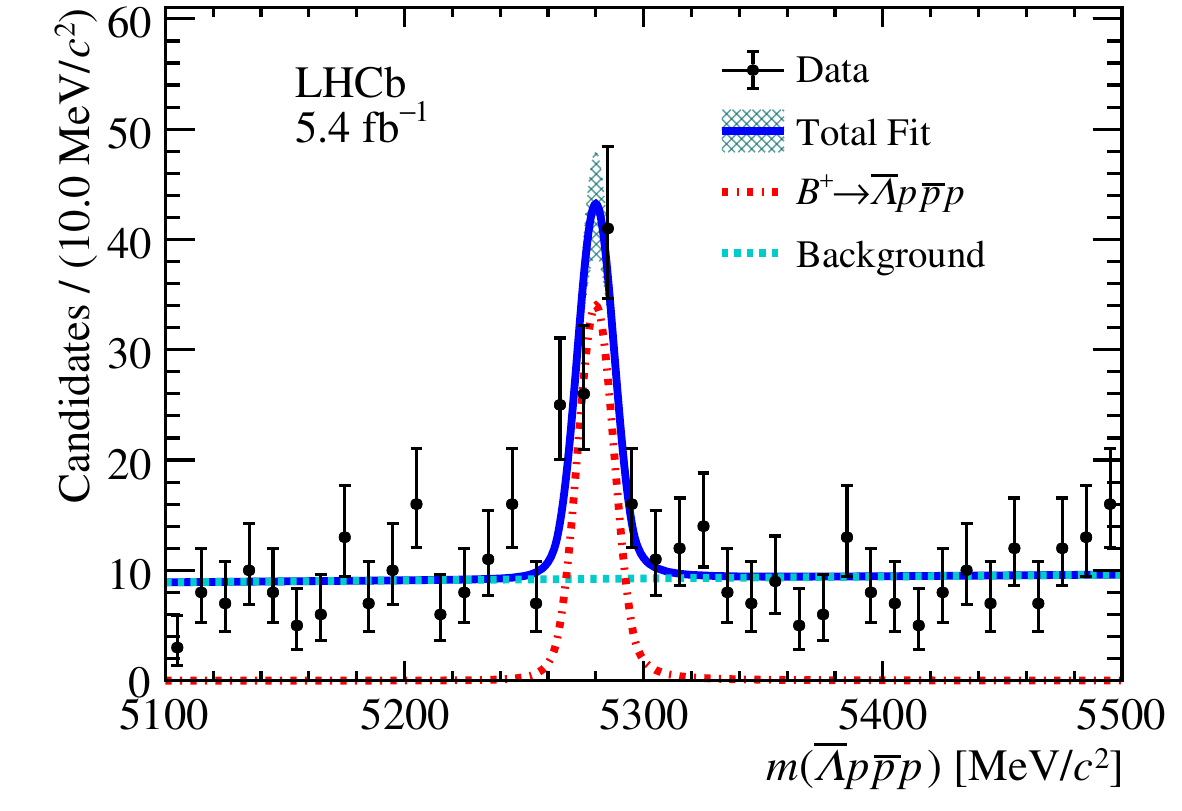}
    \includegraphics[width=0.45\textwidth]{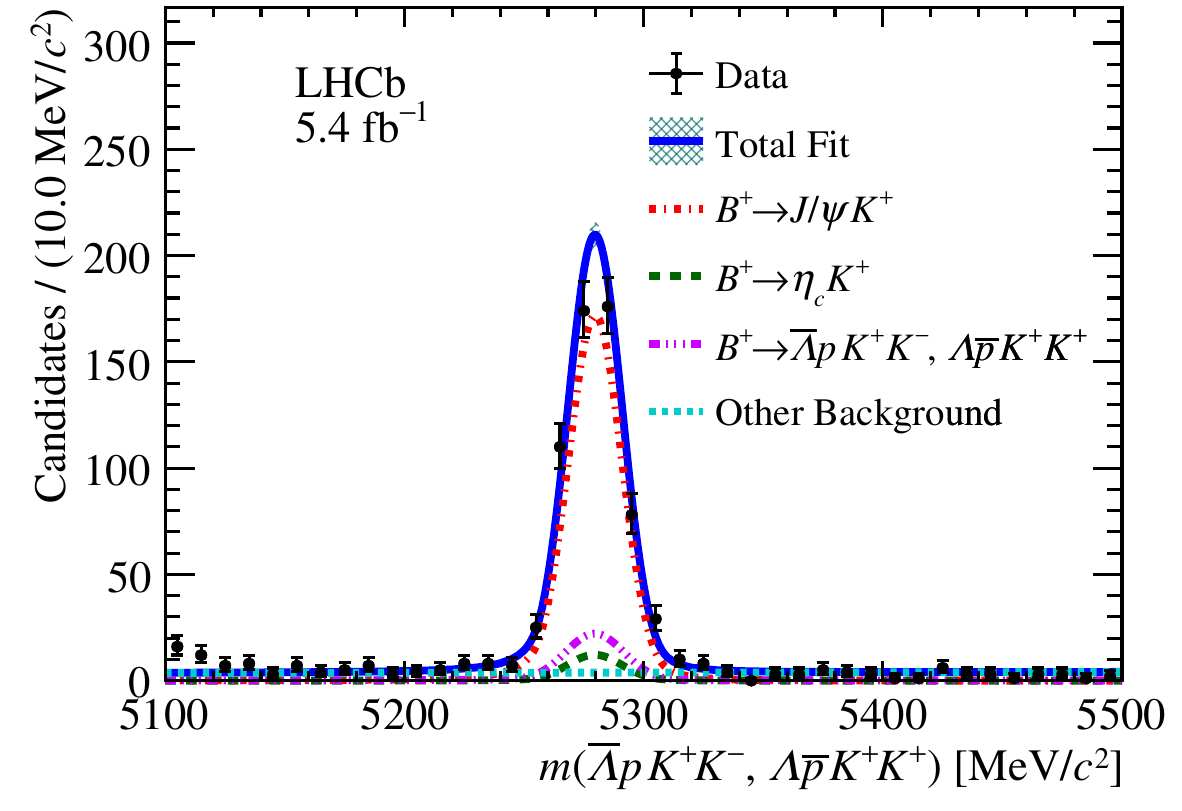}     
    \caption{Invariant-mass distributions of (left) \BuLPbarPPbar with $m(\antiproton\proton)<2850\mevcc$ and (right) \normalization candidates, after combining the long and downstream $\Lz$ categories and requiring all selection criteria described in the text. The fit results are shown together with the fit model components. The hashed cyan bands represent the $\pm \, 1\sigma$ model uncertainties from the covariance matrices of the best-fit parameters.}    
    \label{fig:SignalMasses}
\end{figure}

The invariant-mass distributions $m(\Lbar\proton\antiproton\proton)$ with \mbox{$m(p\bar{p})<2850\mevcc$} and  $m(\Lbar\proton KK)$ are shown in Fig.~\ref{fig:SignalMasses}. 
Because of the different available phase space, the final-state particles in the signal decay have lower momenta, resulting in an improved reconstructed mass resolution. Unbinned maximum-likelihood fits are performed to determine the signal yields.
The signal shapes are described by a modified Gaussian function with tails on both sides of the peak (double-sided Crystal Ball~\cite{Skwarnicki:1986xj}, DSCB) whose values are fixed from simulation. 
The $\Bp$ mass and width are allowed to vary independently. 
The background shapes are described by linear functions with coefficients that are free to vary. 
The \Bp signal yield obtained from the fit to the $m(\Lbar\proton\antiproton\proton)$ selected candidates is \mbox{$N(\BuLPbarPPbar)=78\pm12$} events, accounting for statistical uncertainties only. %% mds This corresponds to a significance greater than five standard deviations.

The absolute branching fraction \mbox{$\mathcal{B}(\BuLPbarPPbar)$} is determined from a simultaneous unbinned maximum-likelihood fit to the signal and the normalization channels.
In the normalization channel \normalizationcc, decays of \mbox{$\Bp \to \etac(\to \Lbar\proton \Km +  c.c.)\Kp$}, 
and nonresonant \mbox{$\Bp \to \Lbar\proton \Kp\Km$} and 
\mbox{$\Bp \to \Lz\antiproton \Kp\Kp$} 
can produce physics backgrounds. 
To account for these, \mbox{\normalizationcc} candidates are divided into three categories based on the invariant mass $m(\Lbar \proton \Km)$: (i) candidates with \mbox{$|m(\Lbar \proton \Km)-M_{\jpsi}|<30\mevcc$} are considered from \normalization~decay, (ii) candidates with \mbox{$|m(\Lbar \proton \Km)-M_{\etac}|<30\mevcc$} are considered from \mbox{$\Bp \to \etac(\to \Lbar\proton \Km)\Kp$} decay,\footnote{The $m(\Lbar \proton \Km)$ observed width is approximately $13.8\mevcc$ for $\eta_c\to\Lbar \proton \Km$ and $4.2\mevcc$ for $\jpsi\to\Lbar \proton \Km$.} and (iii) candidates with \mbox{$3200<m(\Lbar \proton \Km)<3300\mevcc$} are considered from charmless \mbox{$\Bp \to \Lbar\proton \Kp\Km$} and \mbox{$\Bp \to \Lz\antiproton \Kp\Kp$} decays. For each decay mode, the invariant-mass distributions
$m(\Lbar\proton \Kp\Km)$ and $m(\Lz\antiproton \Kp\Kp)$ are modeled using DSCB functions.  
The background yields of \mbox{$\Bp \to \etac(\to \Lbar\proton \Km)\Kp$}, \mbox{$\Bp \to \Lbar\proton \Kp\Km$}, and \mbox{$\Bp \to \Lz\antiproton \Kp\Kp$} decays in category (i) are extrapolated from the $\Bp$ signal yields in categories (ii) and (iii). 
Individual contributions within the $\jpsi$ mass window
are estimated using extrapolations from simulated data.

\begin{figure}[htb]
    \centering
        \includegraphics[width=0.32\linewidth]{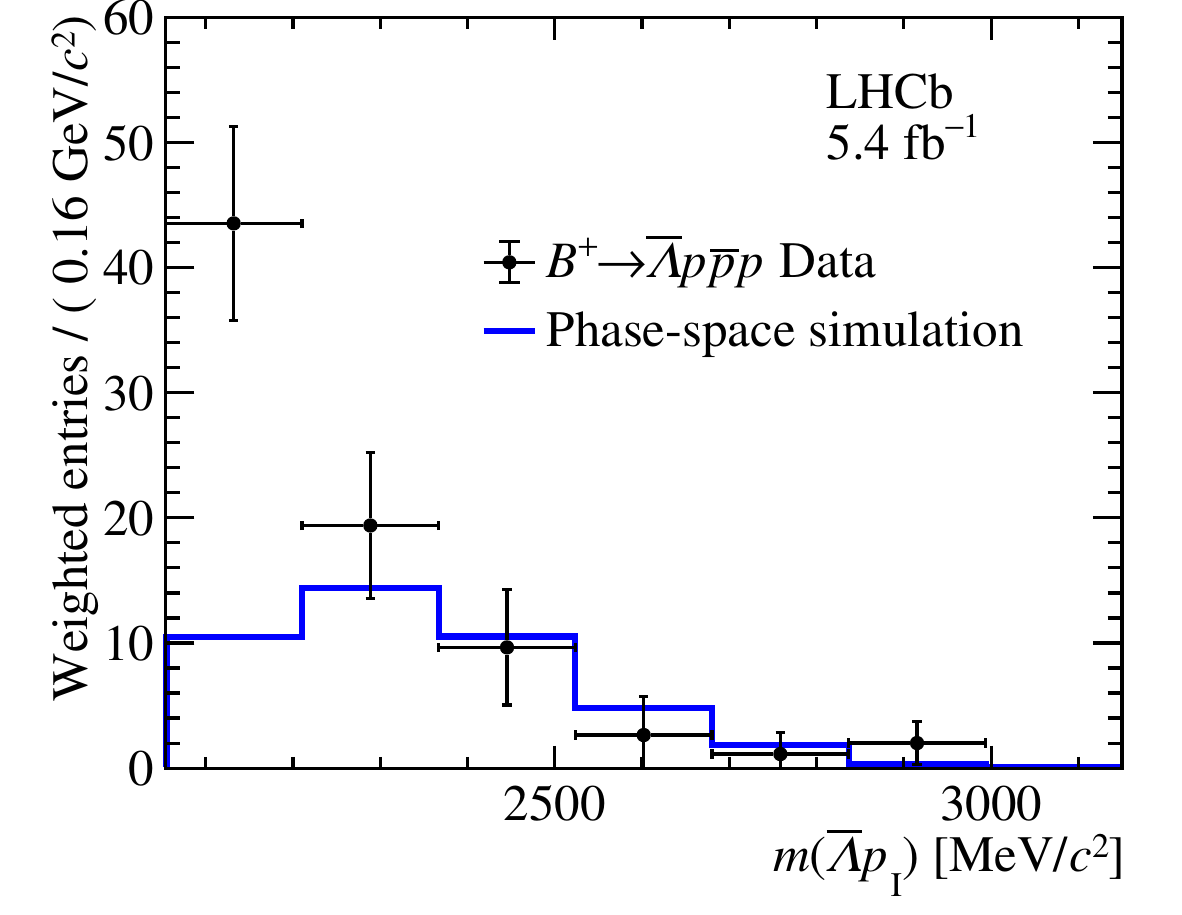}
        \includegraphics[width=0.32\linewidth]{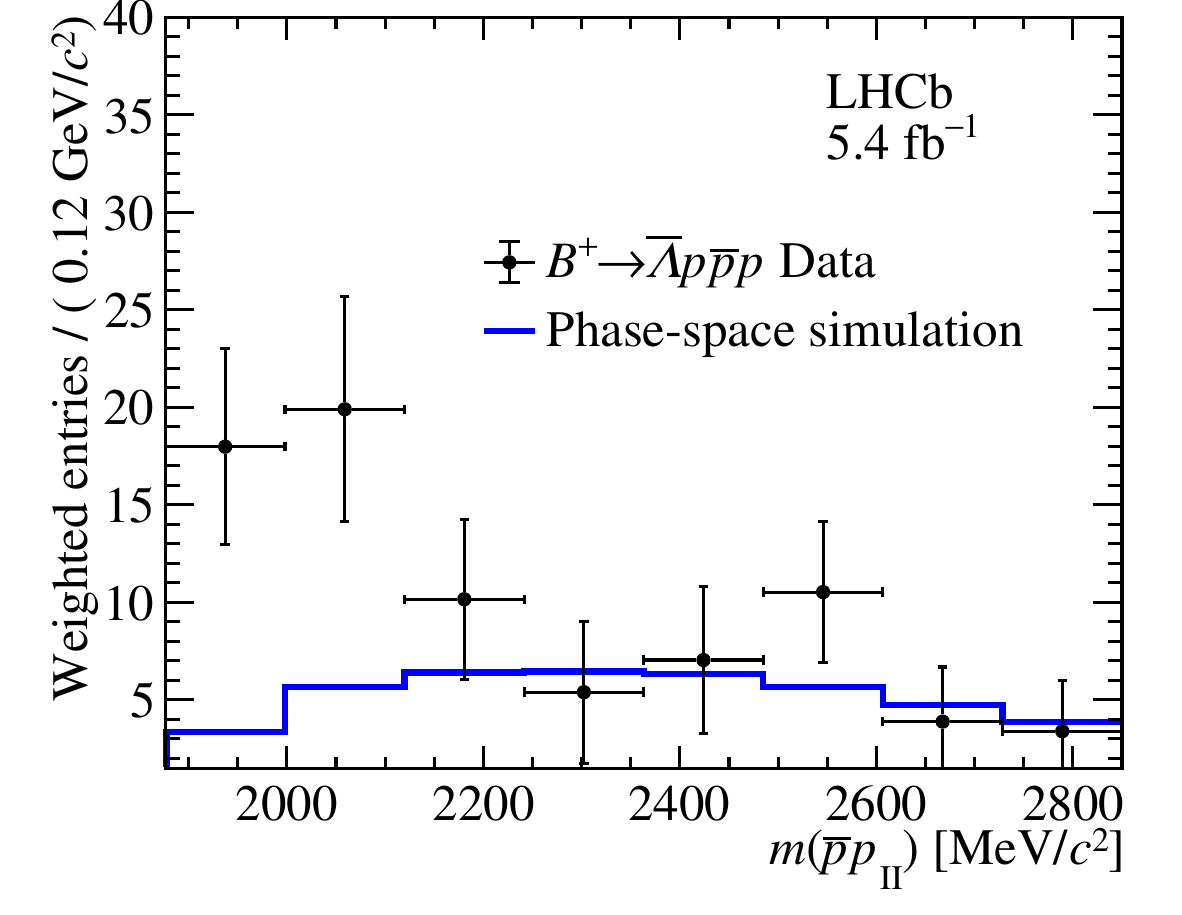}
        \includegraphics[width=0.32\linewidth]{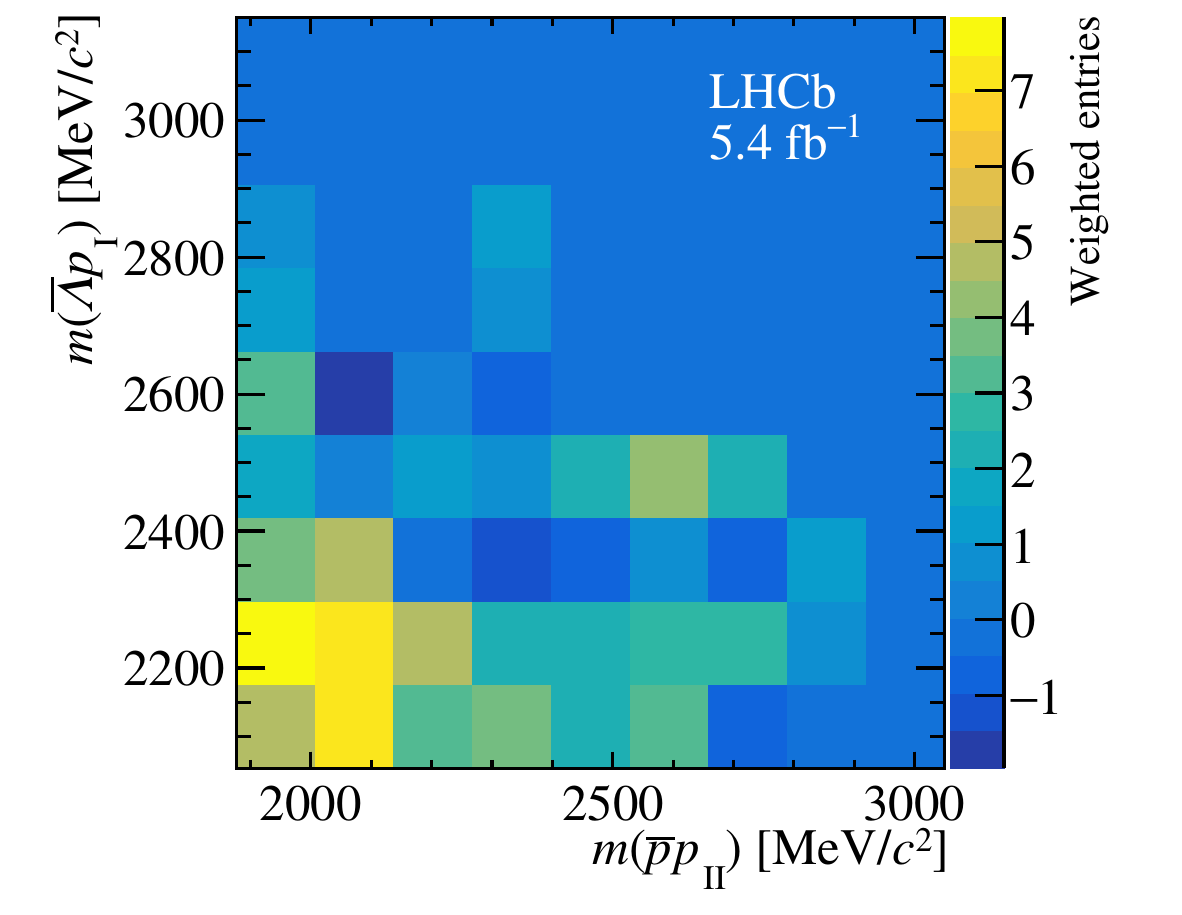}
    \caption{Distributions of background-subtracted (left) $m(\Lbar\proton_{\rm I})$  and (middle) $m(\antiproton\proton_{\rm II})$ for \mbox{\BuLPbarPPbar} events in data compared with phase-space simulation, where the proton $\proton_{\rm I}$ yields the lower invariant mass $m(\Lbar\proton)$ with $\Lbar$ baryon. (Right) Two-dimensional distribution of $m(\Lbar\proton_{\rm I})$ and $m(\antiproton\proton_{\rm II})$ in data.}
    \label{fig:massSpectra}
\end{figure}

The background-subtracted invariant-mass spectra of $\Lbar\proton$ and $\antiproton\proton$ pairs, obtained using the \sPlot technique in the $m(\Lbar\proton\antiproton\proton)$ dimension~\cite{Pivk:2004ty}, are shown in Fig.~\ref{fig:massSpectra}. The two identical protons in the final state are distinguished based on the invariant masses formed with the $\Lbar$ baryon.
The proton that yields the lower invariant mass $m(\Lbar\proton)$ is labeled as $\proton_{\rm I}$. 
The other proton, $\proton_{\rm II}$, is paired with the antiproton to form the second baryon–antibaryon combination. 
A double threshold enhancement is observed, indicated by the clear accumulation of events in the lower left-hand corner of the weighted $(m(\Lbar\proton_{\rm I}),\ m(\antiproton \proton_{\rm II}))$ distribution, as well as by the peaks near the kinematic thresholds in the $m(\Lbar\proton_{\rm I})$ and $m(\antiproton\proton_{\rm II})$ spectra.
Similar results, obtained by pairing the proton and antiproton with the lower invariant mass, are shown in the End Matter.

\begin{figure}[b]
    \centering
    \includegraphics[width=0.45\textwidth]{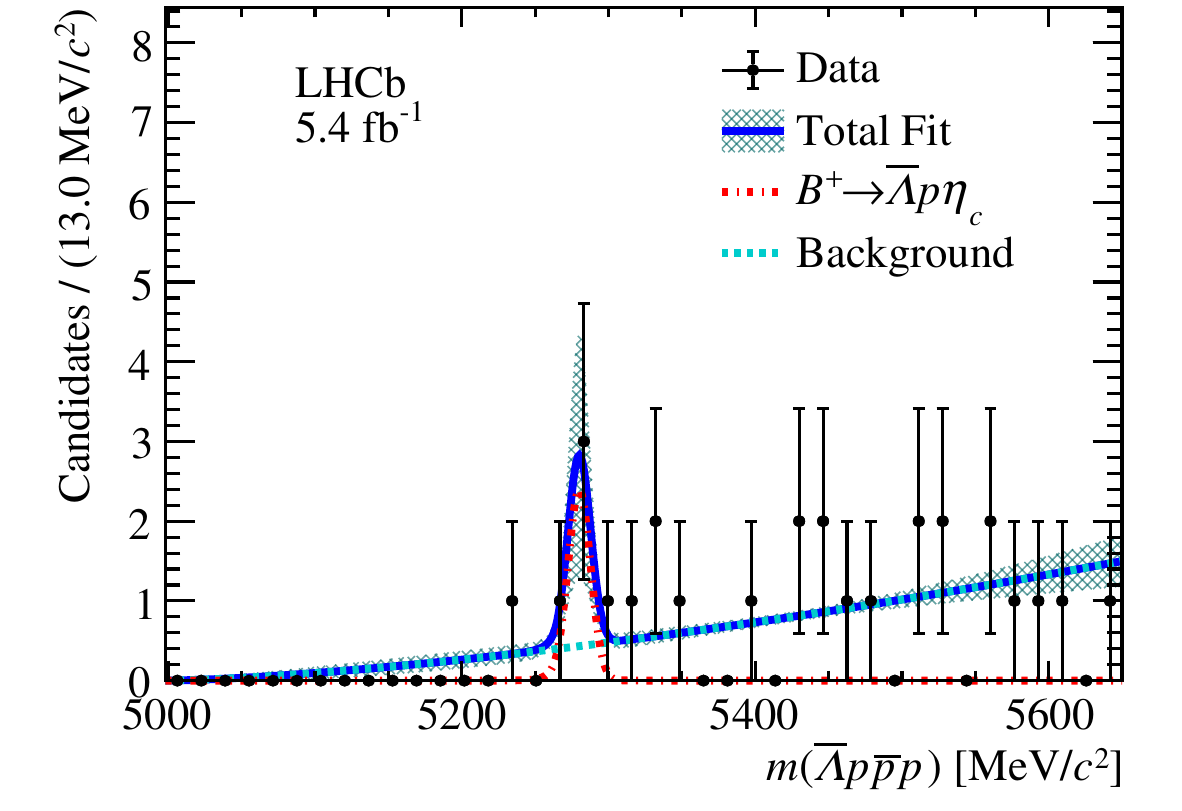}
    \includegraphics[width=0.45\textwidth]{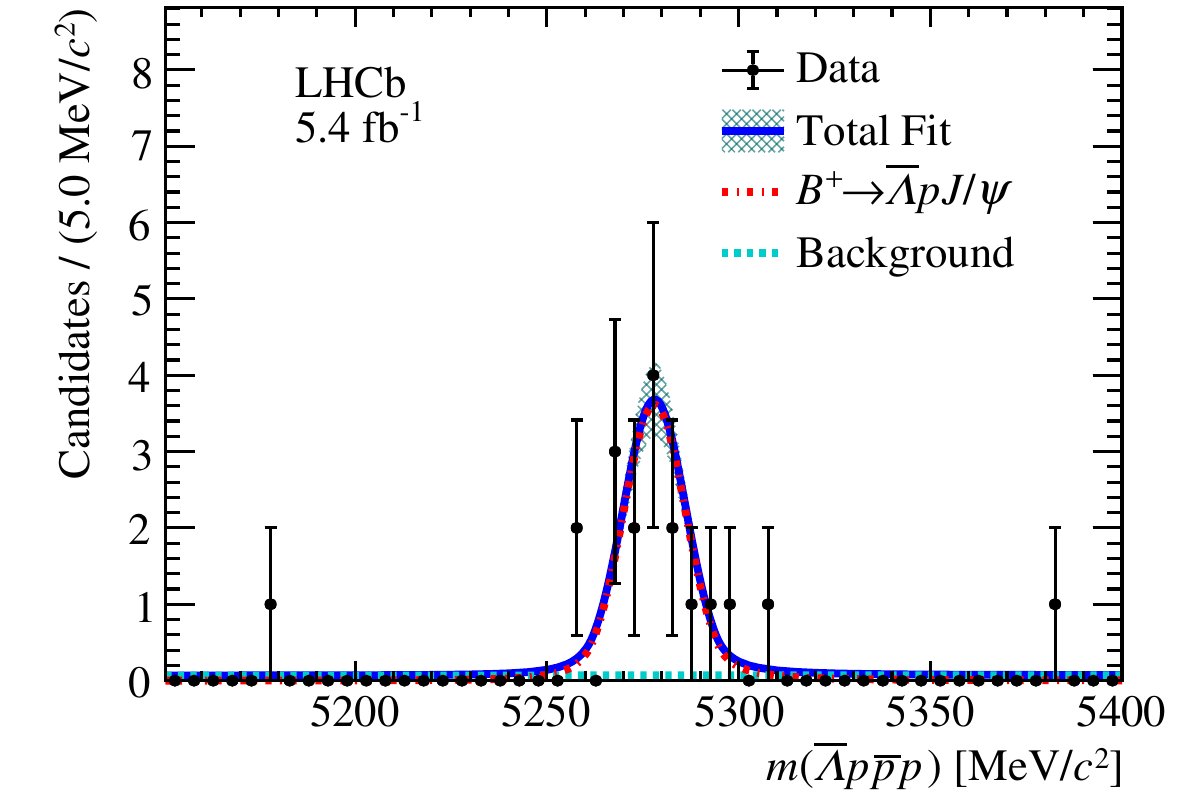}
    \caption{Invariant-mass distribution of (left) \BuLPbarEtac~candidates with \mbox{$|m(\antiproton\proton)-M_{\etac}|<40\mevcc$} and (right) \mbox{$\Bp\to \Lbar\proton\jpsi(\to \antiproton\proton)$} candidates with \mbox{$|m(\antiproton\proton)-M_{\jpsi}|<30\mevcc$}, after combining the long and downstream $\Lz$ categories and requiring all selection criteria described in the text. The fit results are shown together with the fit model components. The hashed cyan bands represent the $1\sigma$ model uncertainty on the covariance matrix of the best-fit parameters.}
    \label{fig:EtacSignalMasses}
\end{figure}

The $m(\Lbar\proton\antiproton\proton)$ distribution  of candidates with 
\mbox{$|m(\antiproton\proton) - M_{\etac}| < 40\mevcc$} is shown in Fig.~\ref{fig:EtacSignalMasses}. The signal shape is modeled by a DSCB function and the background by a threshold function of the form $(x-x_0)^{\beta}$, where \mbox{$x_0\equiv M_{\Lz}+M_{\eta_c}+M_{p}$} and $\beta$ is a free parameter of the fit. The yield in this region is $N(\BuLPbarEtac)=2.9 \pm 1.9$ events. 
%% mds (statistical uncertainty only). 
The yield in the region $|m(\antiproton\proton) - M_{\jpsi}| < 30\mevcc$ is  \mbox{$N(\Bp\to \Lbar\proton\jpsi(\to \antiproton\proton))=16.4\pm4.2$} events. This is consistent with estimates from the known branching fractions 
 $ {\cal B}(\Bp\to \Lbar\proton\jpsi) $ and
$ {\cal B}(\jpsi \to p \bar p )$~\cite{PDG2024} and the number of 
events observed in the normalization channel.
 
%% ==========================
%%   Systematic uncertainties
%% ==========================

The dominant sources of systematic uncertainty affecting the branching fraction measurements are listed in Table~\ref{tab:summary_syst_BF}, and include those associated with the fit models, efficiencies, PID performance, and the branching fraction of the normalization channel. 

Three sources of associated systematic uncertainties
are estimated from simulation. 
First are the uncertainties on the nominal efficiencies due to the sizes of the simulated samples.
Second are the uncertainties due to the weights used for correcting kinematic differences between simulation and data, assessed by recalculating the branching fractions without applying the weights. 
The resulting deviation is taken as a conservative estimate of the associated uncertainty.
Third are uncertainties on the tracking efficiencies, 
due to the detector material budget distribution or to the particle interaction cross-sections, 
that lead to uncertainties on the efficiency ratios. 
Differences between PID efficiencies in data and simulation are corrected using a combination of simulated and data calibration samples. 
%% mds The corrections depend on both the sizes of the simulation samples and the procedure used to weight the data calibration samples to match the kinematic distributions of the signals.  
The signal shapes are varied in two ways: the DSCB function parameters are varied within their uncertainties, and the signal shape is modified to include a Gaussian function together with the DSCB function. For the \BuLPbarPPbar decay, the background model is replaced by a second-order polynomial function. For the \BuLPbarEtac decay, the threshold background shape is replaced by an exponential function. Concerning the correction of decay models for signal and normalization channels in simulation, the parameters that affect the performance of the weighting algorithm are varied individually, and the total efficiencies are recalculated using the alternative sets of weights, with the largest deviation taken as a systematic uncertainty. For $\BuLPbarPPbar$ decay, an alternative correction based on the decay model shown in Fig.~\ref{Mat:Plot_MassSpectra} is implemented and the resulting deviation is taken as an additional contribution to the systematic uncertainty associated with the signal model correction. For the $\BuLPbarEtac$ decay, the dominant systematic uncertainties relate to the signal shape and the potential contribution from charmless $\BuLPbarPPbar$ decay, which is estimated by extrapolating the observed yield with \mbox{$m(\antiproton\proton)<2850\mevcc$} into the $ \etac $ mass window.

%% ------------------------------
%% Systematic uncertainties table (relative to statistical uncertainties)
%% ------------------------------
\begin{table}[t]
\caption{Summary of relative systematic uncertainties on (A) $\BF( \BuLPbarPPbar )$ and (B) \mbox{$\BF( \BuLPbarEtac )$}, in percent. The total systematic uncertainty is computed as the %square root of the sum of squared 
sum in quadrature of
individual contributions. There is no model correction for $\BuLPbarEtac$ decay due to the small size of this component. The last row indicates the relative statistical uncertainty.}
\label{tab:summary_syst_BF}
\begin{center}
\begin{tabular}{l c c}
\toprule 
Systematic source   & A & B\\ 
\midrule 
Simulated sample size	    &            $0.6$   & $\phantom{1}0.6$ \\
Kinematic correction        &            $1.4$   & $\phantom{1}1.4$\\
Tracking 					&            $1.2$   & $\phantom{1}1.2$\\ 
PID 						&            $3.1$   & $\phantom{1}2.0$ \\ 
Signal shape 				&            $2.9$   & $31.1$ \\ 
Background shape 			&            $1.2$   &  $\phantom{1}2.4$ \\
Model correction (signal)   &            $2.7$   & /        \\ 
Model correction (normalization) &       $1.6$   & $\phantom{1}1.6$          \\
Nonresonant                 &               /    & $22.9$            \\ \midrule 
Total systematic            &            $5.8$   & $38.8$ \\ \midrule 
Statistical uncertainty     &            $16.4$  & $65.5$ \\ 
\bottomrule 
\end{tabular} 
\end{center} 
\end{table}

%%%%%%%%%%%%%%%
%%%%% ACP %%%%%
%%%%%%%%%%%%%%%
% https://doi.org/10.1016/j.physletb.2018.10.039
The \CP asymmetry of the $\BuLPbarPPbar$  decay rate integrated over the phase space is measured by fitting the $\Bp$ and $\Bm$ samples separately, as shown in the End Matter, using the same procedure as for the branching fraction measurement. 
 The yields vary independently, while the shape parameters are shared between the $\Bp$ and $\Bm$ candidates. 
 The raw asymmetry ($A_{\rm raw}$) between the \Bp and \Bm signal yields is found to be  $(5.6\pm15.5)\%$. To obtain the \CP asymmetry, this must be corrected as per
\begin{equation}
\mathcal{A}_{\CP}=A_{\rm raw}-A_{\rm prod}-A_{\rm det},\nonumber
\end{equation}
where $A_{\rm prod}$ is the production asymmetry of $B^\pm$ mesons and $A_{\rm det}$ is the sum of the instrumental asymmetries in detection ($A_{\rm D}$), PID ($A_{\rm PID}$), and trigger ($A_{\rm trigger}$) of particles and antiparticles. 
 
The production asymmetry has been measured by LHCb to be $ (-0.7 \pm 0.3)\% $~\cite{LHCb-PAPER-2021-049}. The trigger asymmetry arises from differences in trigger efficiencies between charge-conjugated final states. The efficiency for a charged hadron to be responsible for the affirmative decision of the hardware trigger is determined as a function of \pt, separately for positively and negatively charged particles~\cite{LHCb-DP-2012-004}. The asymmetry introduced by the hardware trigger for candidates is determined to be $A_{\rm trigger}^{\Lbar\proton\antiproton\proton} = (0.03 \pm 0.18)\%$. The proton detection asymmetry is evaluated as a function of momentum using the method developed in Ref.~\cite{LHCb-PAPER-2021-016}, and is measured to be $A_{\rm D}^{\proton\antiproton\proton}=(1.15\pm0.66)\%$. The detection asymmetry for the $\Lz$ baryon is estimated by parameterizing the detection asymmetries of the proton and pion as functions of their momentum and, in the case of the pion, also pseudorapidity~\cite{LHCb-PAPER-2021-016,LHCb-PAPER-2021-009}, and is measured to be $A_{\rm D}^{\Lbar}=(-1.49\pm0.26)\%$. The PID asymmetry is evaluated using calibration samples that account for differences in PID efficiency between positively and negatively charged tracks and is determined to be $A_{\rm PID}^{\proton\antiproton\proton}=(0.18 \pm 1.50)\%$. The PID asymmetry from the $\Lz$ baryon is negligible due to its loose selection criteria.

The estimation of model-related systematic uncertainties in the determination of $A_{\rm raw}$ is performed using the same procedure as for the branching fraction measurement. The differences between the raw asymmetries determined by the two sets of fits are added in quadrature, and the resulting value is taken as a systematic uncertainty.
Altogether with the uncertainties on external asymmetries, the \CP asymmetry is measured to be $ \mathcal{A}_{\CP} = (5.4 \pm 15.6 \pm 2.4) \% $.

%% ============================
%%   Discussion and conclusions
%% ============================

In summary, a new charmless and purely baryonic decay mode of the \Bp meson, \BuLPbarPPbar, is observed. %
% using \pp collision data collected by the \lhcb experiment over the period 2016--2018 at $\sqrt{s}=13$\tev and corresponding to an integrated luminosity of approximately 5.4\invfb.
The signal yield is measured to be \mbox{$N(\BuLPbarPPbar) = 78\pm12$} events.
% , corresponding to a significance greater than five standard deviations.
The branching fraction relative to the normalization channel is determined to be \mbox{$\BF(\BuLPbarPPbar)/\BF(\normalization) = (0.245\pm 0.040\pm 0.014)$}, where the uncertainties are statistical and systematic, respectively. The branching fraction is measured to be 
\begin{equation*}
    \BF(\BuLPbarPPbar) = (2.15\pm0.35\pm0.12\pm0.28) \times 10^{-7}, 
\end{equation*}where the first uncertainty is statistical, the second systematic, and the third arises from the external branching fraction of the normalization channel. 
This result is lower than, but in reasonable agreement with, a SM prediction of \mbox{$\BF(\BuLPbarPPbar) = (7.4^{+0.6}_{-0.2} \pm 0.03 ^{+3.6}_{-2.6}) \times 10^{-7}$~\cite{Hsiao:2022tfj}}, where the uncertainty arises from nonfactorizable QCD corrections, the CKM matrix-element inputs, and  form‐factor modeling. The \CP asymmetry is measured to be \mbox{$\mathcal{A}_{\CP}=(5.4\pm 15.6\pm 2.4)\%$}, with the first uncertainty statistical and the second systematic. The background-subtracted mass spectra exhibit a clear double-threshold enhancement in both baryon–antibaryon invariant-mass distributions close to their thresholds, which plays an enhancing factor \cite{Hsiao:2022tfj} in the \BuLPbarPPbar decay and may reveal resonant structures such as the $X(1835)$ baryonium bound state~\cite{BESIII:2011aa}.

In the region defined by \mbox{$|m(\antiproton\proton)-M_{\eta_c}|<40\mevcc$}, the yield of \BuLPbarEtac is \mbox{$N(\BuLPbarEtac)=2.9 \pm 1.9$} events. 
The corresponding branching fraction is 
\begin{equation*}
\BF(\BuLPbarEtac) = 
    (8.3\pm5.4\pm3.2\pm1.1)\times 10^{-9},
\end{equation*}
where the first uncertainty is statistical, the second systematic, and the third arises from the external branching fraction of the normalization channel. The $90\%$ confidence level upper limit on the branching fraction is \mbox{$\mathcal{B}(\BuLPbarEtac)<2.1\times 10^{-8}$}.

% The $90\%$ confidence level upper limit, obtained by requiring that the change in the profile likelihood is less than 2.71, is \mbox{$\mathcal{B}(\BuLPbarEtac)<2.1\times 10^{-8}$}.

%

The measurement of $\mathcal{B}(\BuLPbarPPbar)$ constrains models of perturbative QCD effects in multibody baryonic final states, particularly in regions where large theoretical uncertainties remain~\cite{Hsiao:2022tfj}. 
The threshold enhancements manifested in this study highlight the rich dynamics of baryonic \B decays and should stimulate further theoretical studies.

%TC:ignore
\section*{Acknowledgements}
%
% These Acknowledgements valid from 3-May-2019
%
\noindent We express our gratitude to our colleagues in the CERN
accelerator departments for the excellent performance of the LHC. We
thank the technical and administrative staff at the LHCb
institutes.
We acknowledge support from CERN and from the national agencies:
ARC (Australia);
CAPES, CNPq, FAPERJ and FINEP (Brazil); 
MOST and NSFC (China); 
CNRS/IN2P3 (France); 
BMBF, DFG and MPG (Germany); 
INFN (Italy); 
NWO (Netherlands); 
MNiSW and NCN (Poland); 
MCID/IFA (Romania); 
%MSHE (Russia); 
MICIU and AEI (Spain);
SNSF and SER (Switzerland); 
NASU (Ukraine); 
STFC (United Kingdom); 
DOE NP and NSF (USA).
%%%%%%%%%%%%%%%%%%%%%%%%%%%%%%%%%%%%%%%%%%%%%
We acknowledge the computing resources that are provided by ARDC (Australia), 
CBPF (Brazil),
CERN, 
IHEP and LZU (China),
IN2P3 (France), 
KIT and DESY (Germany), 
INFN (Italy), 
SURF (Netherlands),
Polish WLCG (Poland),
IFIN-HH (Romania), 
%RRCKI and Yandex LLC (Russia), 
PIC (Spain), CSCS (Switzerland), 
and GridPP (United Kingdom).
%%%%%%%%%%%%%%%%%%%%%%%%%%%%%%%%%%%%%%%%%%
We are indebted to the communities behind the multiple open-source
software packages on which we depend.
%%%%%%%%%%%%%%%%%%%%%%%%%%%%%%%%%%%%%%%%%%
Individual groups or members have received support from
%ARC and ARDC (Australia); % moved to national 16/01
Key Research Program of Frontier Sciences of CAS, CAS PIFI, CAS CCEPP, 
Fundamental Research Funds for the Central Universities,  and Sci.\ \& Tech.\ Program of Guangzhou (China);
Minciencias (Colombia);
EPLANET, Marie Sk\l{}odowska-Curie Actions, ERC and NextGenerationEU (European Union);
A*MIDEX, ANR, IPhU and Labex P2IO, and R\'{e}gion Auvergne-Rh\^{o}ne-Alpes (France);
%RFBR, RSF and Yandex LLC (Russia);
Alexander-von-Humboldt Foundation (Germany);
ICSC (Italy); 
%GVA, XuntaGal, GENCAT, Inditex, InTalent and Prog.~Atracci\'on Talento, CM (Spain);
Severo Ochoa and Mar\'ia de Maeztu Units of Excellence, GVA, XuntaGal, GENCAT, InTalent-Inditex and Prog.~Atracci\'on Talento CM (Spain);
SRC (Sweden);
the Leverhulme Trust, the Royal Society and UKRI (United Kingdom).
\\

\emph{Data availability}—The data that support the findings of this article are openly available~\cite{cdslink}.

\begingroup
  \makeatletter
    \let\addcontentsline\@gobblethree
  \makeatother
  \clearpage
\section*{End Matter}

\section{Additional baryon–antibaryon mass spectra}
\label{Mat:MassSpectra}

Another study of baryon–antibaryon invariant-mass spectra in the $\BuLPbarPPbar$ decay is shown in Fig.~\ref{Mat:Plot_MassSpectra}. 
The two identical protons in the final state are distinguished based on the invariant mass formed with the antiproton rather than with the $\Lbar$ baryon. The $\antiproton\proton$ combination that yields the lower invariant mass is labeled as $\antiproton\proton_{\rm 1}$. The remaining proton, $\proton_{\rm 2}$, is then paired with the $\Lbar$ baryon to form the second baryon–antibaryon combination. 

\begin{figure}[hbt]
    \centering
    \includegraphics[width=0.32\linewidth]{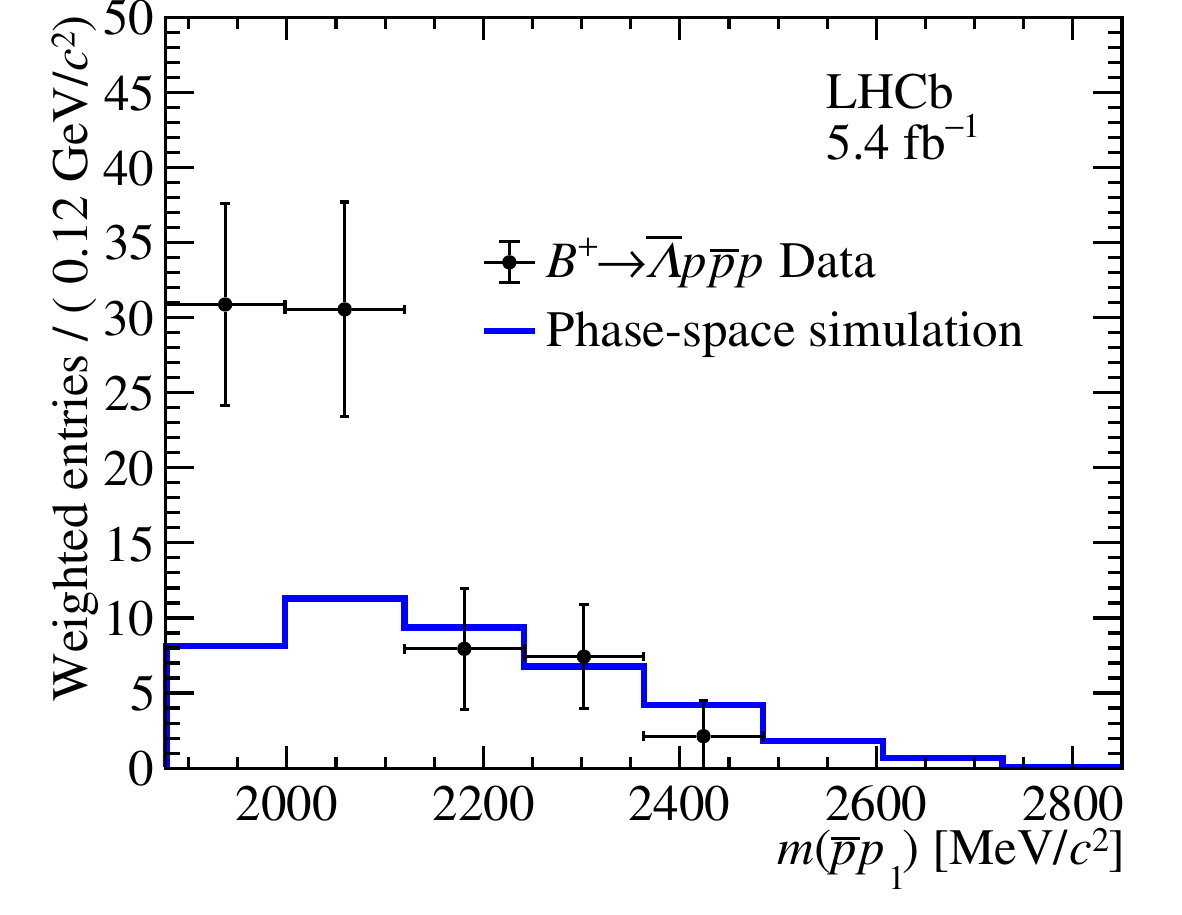}
    \includegraphics[width=0.32\linewidth]{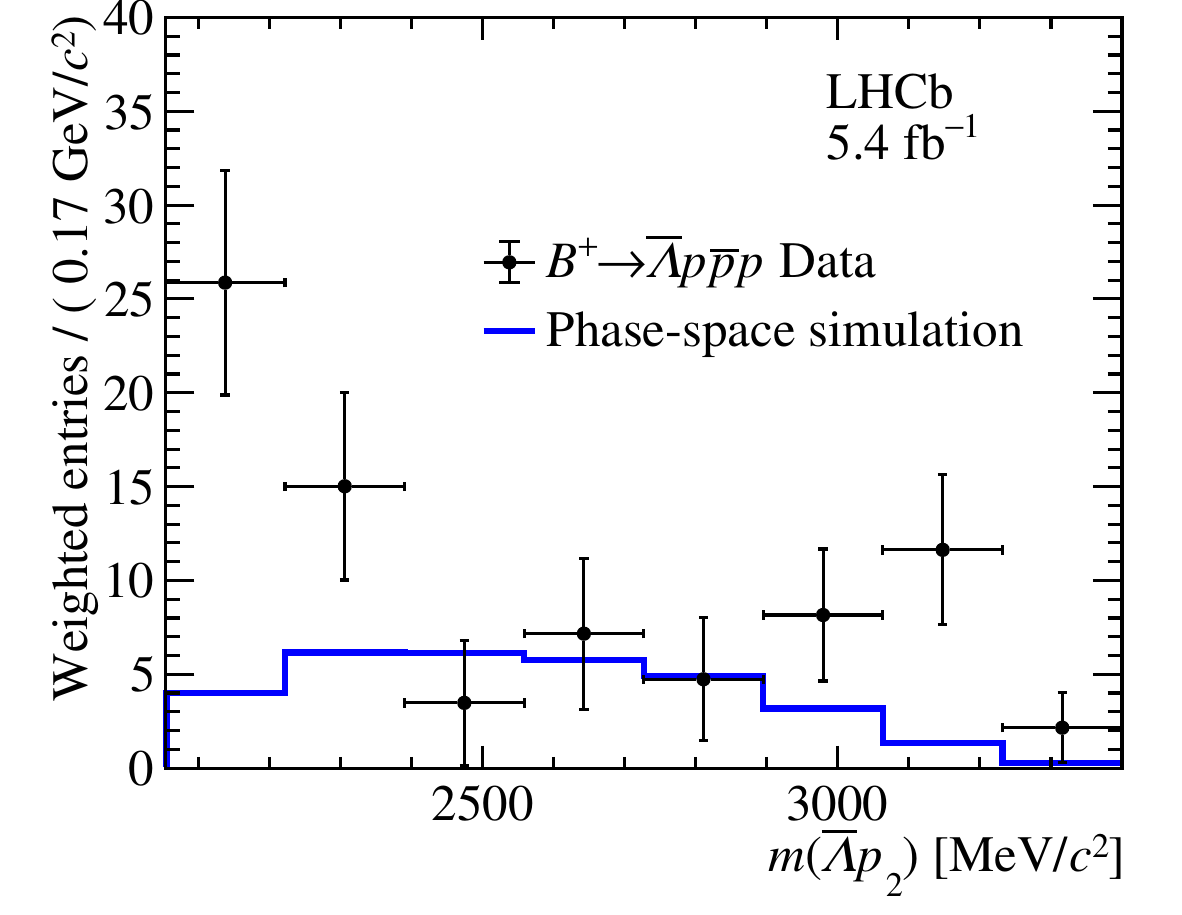}
    \includegraphics[width=0.32\linewidth]{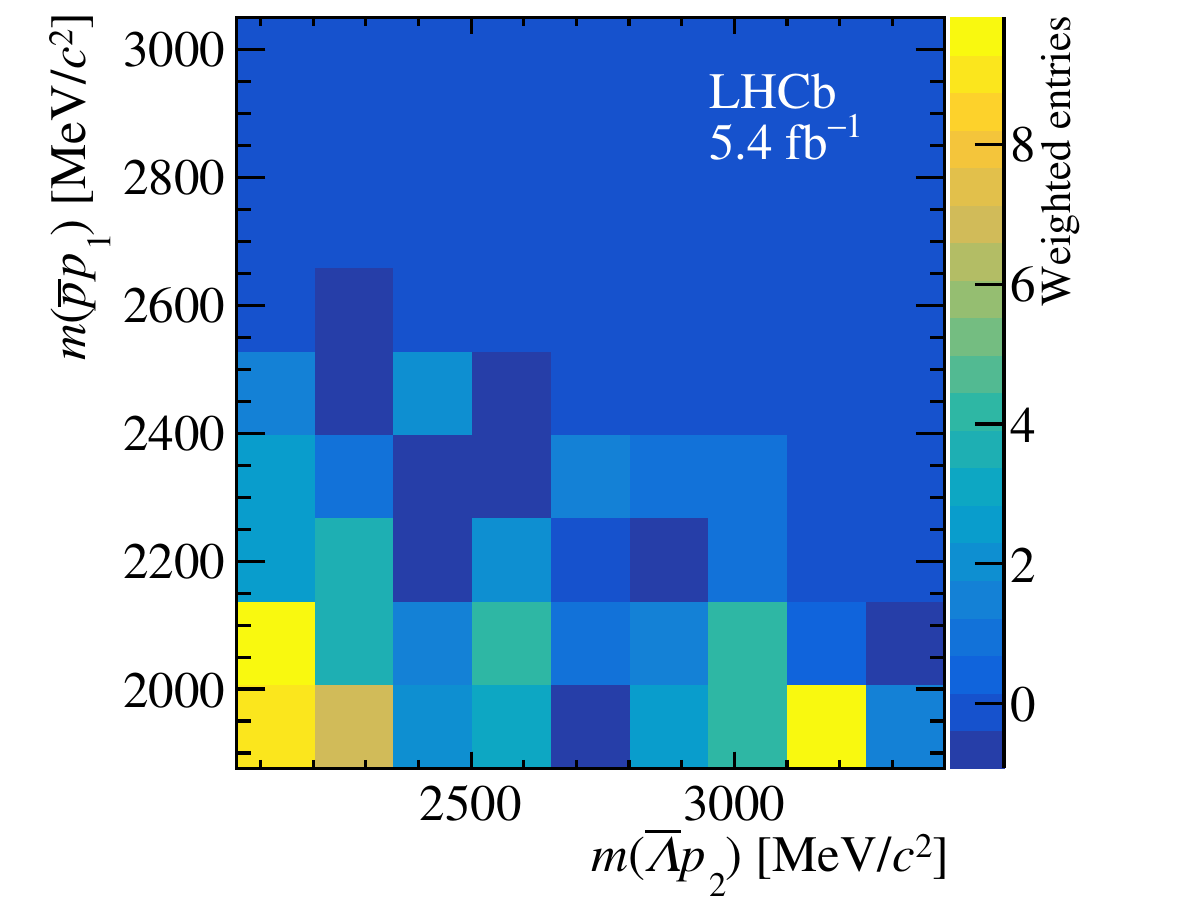}
    \caption{Distributions of background-subtracted (left) $m(\antiproton\proton_1)$  and (middle) $m(\Lbar\proton_{2})$ for \mbox{\BuLPbarPPbar} events in data compared with phase-space simulation, where the proton $\proton_{\rm 1}$ yields the lower invariant mass  with antiproton. (Right) Two-dimensional distribution of $m(\antiproton\proton_1)$ and $m(\Lbar\proton_{2})$ in data. }
    \label{Mat:Plot_MassSpectra}
\end{figure}

As expected due to the definition of the pairs, the enhancement here is 
greater for $ m ( \antiproton\proton ) $ than in Fig.~\ref{fig:massSpectra} and less for
$  m ( \Lbar \proton ) $.
Qualitatively, however, the conclusion is the same and enhancements are observed in the lower left-hand corner of the scatter plot and in both projections.

\section{Fit to different charged \boldmath{$\B$} samples}
\label{Mat:CPFIT}
Figure~\ref{fig:cpvfit} shows the fits to the $\Bu \to \Lbar \proton\antiproton\proton$ and $\Bm \to \Lz \antiproton\proton\antiproton$ samples for the measurement of $\mathcal{A}_{\CP}$.

\begin{figure}[tb!]
    \centering
    \includegraphics[width=0.45\linewidth]{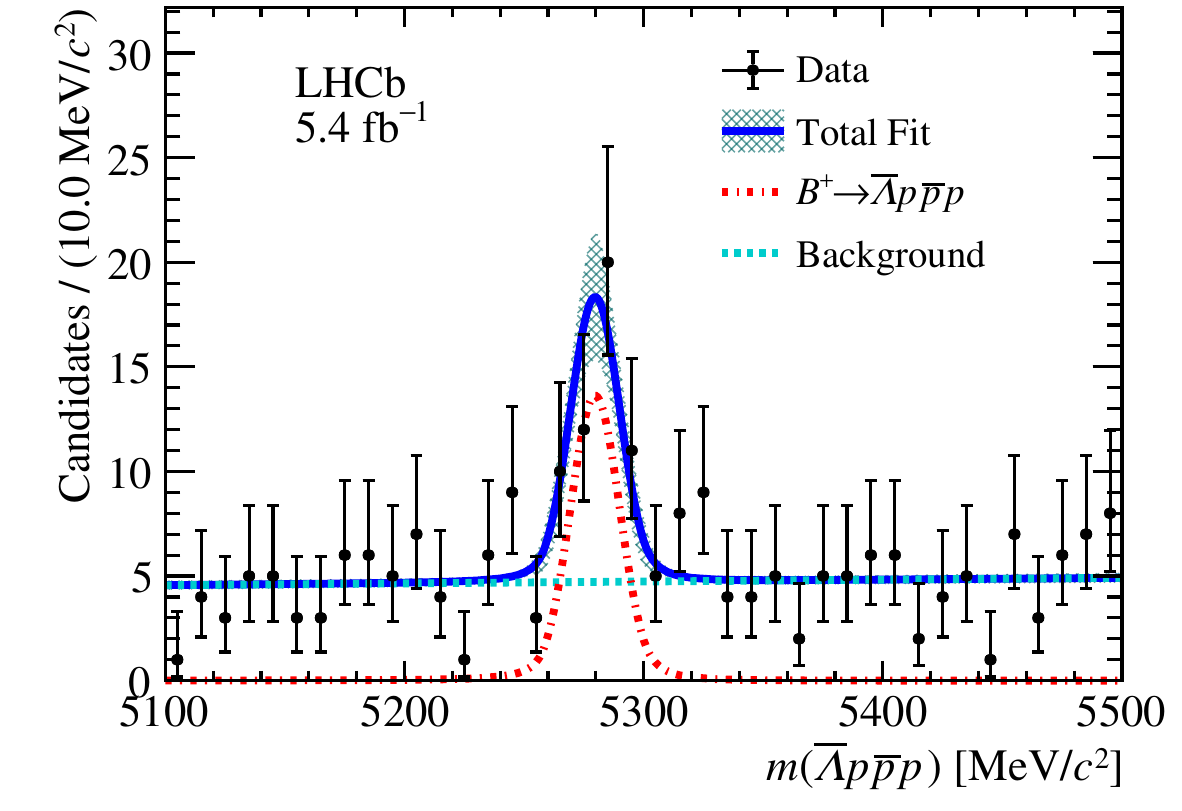}
    \includegraphics[width=0.45\linewidth]{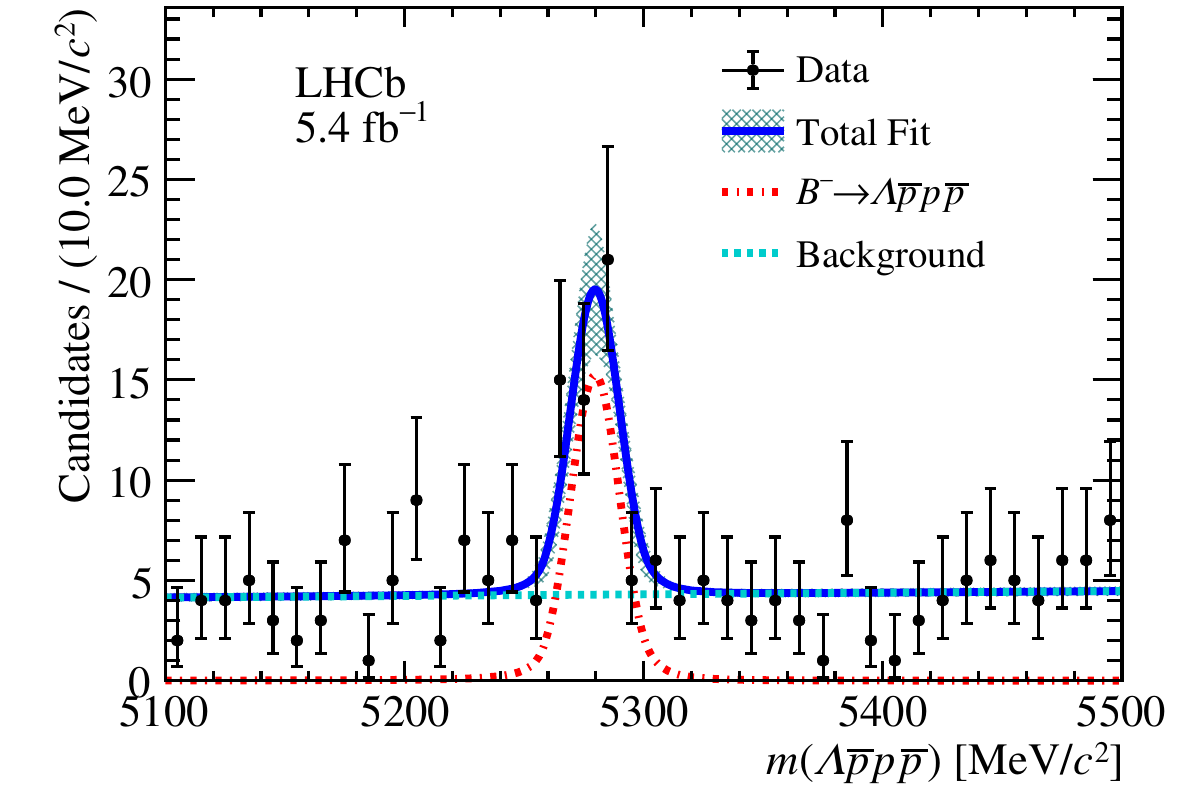}
    \caption{Invariant-mass distribution of (left) $\Bp\to \Lbar \proton\antiproton\proton$  and  (right) $\Bm\to \Lz\antiproton\proton\antiproton$ candidates, after combining the long and downstream $\Lz$ categories and requiring all selection criteria described in the text. The fit results for these samples are shown together with the fit model components. The hashed cyan bands represent the $1\sigma$ model uncertainty on the covariance matrix of the best-fit parameters. }
    \label{fig:cpvfit}
\end{figure}

\clearpage
%%%%%%%%%%%%%%%%%%%%%%%%%%%%%%%%%%%%%%%%%%%%%%%%%%%%%%%
  % not put sections into ToC
\endgroup   

\clearpage
\addcontentsline{toc}{section}{References}
\setboolean{inbibliography}{true}
\bibliographystyle{LHCb}
\bibliography{main,LHCb-PAPER,LHCb-CONF,LHCb-DP,LHCb-TDR,standard}

\newpage
% LHCb collaboration author list
% Data extracted on August 21st, 2025 at 4:33am for paper reference LHCb-PAPER-2025-032
\centerline
{\large\bf LHCb collaboration}
\begin
{flushleft}
\small
R.~Aaij$^{38}$\lhcborcid{0000-0003-0533-1952},
A.S.W.~Abdelmotteleb$^{57}$\lhcborcid{0000-0001-7905-0542},
C.~Abellan~Beteta$^{51}$\lhcborcid{0009-0009-0869-6798},
F.~Abudin{\'e}n$^{57}$\lhcborcid{0000-0002-6737-3528},
T.~Ackernley$^{61}$\lhcborcid{0000-0002-5951-3498},
A. A. ~Adefisoye$^{69}$\lhcborcid{0000-0003-2448-1550},
B.~Adeva$^{47}$\lhcborcid{0000-0001-9756-3712},
M.~Adinolfi$^{55}$\lhcborcid{0000-0002-1326-1264},
P.~Adlarson$^{85}$\lhcborcid{0000-0001-6280-3851},
C.~Agapopoulou$^{14}$\lhcborcid{0000-0002-2368-0147}
% ...

{\footnotesize \it

% ...
$^{14}$Universit{\'e} Paris-Saclay, CNRS/IN2P3, IJCLab, Orsay, France\\
% ...
$^{38}$Nikhef National Institute for Subatomic Physics, Amsterdam, Netherlands\\
% ...
$^{47}$Instituto Galego de F{\'\i}sica de Altas Enerx{\'\i}as (IGFAE), Universidade de Santiago de Compostela, Santiago de Compostela, Spain\\
% ...
$^{51}$Physik-Institut, Universit{\"a}t Z{\"u}rich, Z{\"u}rich, Switzerland\\
% ...
$^{55}$H.H. Wills Physics Laboratory, University of Bristol, Bristol, United Kingdom\\
% ...
$^{57}$Department of Physics, University of Warwick, Coventry, United Kingdom\\
% ...
$^{60}$School of Physics and Astronomy, University of Glasgow, Glasgow, United Kingdom\\
$^{61}$Oliver Lodge Laboratory, University of Liverpool, Liverpool, United Kingdom\\
% ...
$^{69}$Syracuse University, Syracuse, NY, United States\\
% ...
$^{85}$Department of Physics and Astronomy, Uppsala University, Uppsala, Sweden, associated to $^{60}$\\
% ...
}
\end{flushleft}
% \input{PRL_Justification}

%TC:endignore
\end{document}